# Title: Flux Jumps up to 17 T in ReBCO Tape Stack Cables and their Suppression with Increased Intertape Spacing

**Authors:** Tushar Garg[1]*, Mike D. Sumption[1], Milan Majoros[1], Edward Collings[1], Jan Jaroszynski[2], Eun-Sang Choi[2]

**Affiliations:**

[1]Center for Superconducting and Magnetic Materials, Materials Science Department, Ohio State University, Columbus, OH, 43210, USA.

[2]National High Magnetic Field Laboratory, Tallahassee, Florida, 32310, USA.

*Corresponding author. Email: garg.206@osu.edu

**Abstract:** The magnetization of ReBCO tape stacks and tape stack cables in high magnetic fields (up to 30 T) are not commonly reported. Here we report magnetization measurements of tape stack cables in magnetic fields up to 30 T at 4.2 K. We observed that flux jumps, commonly relegated to low field regimes for single tapes, persisted up to 17 T in tape stacks, an effect which could have substantial technological relevance for high field applications including fusion devices or accelerator magnets. On the other hand, with the use of small spacers, we could suppress flux jump behavior, in some cases eliminating jumps entirely. Our findings provide critical insights for the optimization of designs for ReBCO cables for high-field applications, including fusion magnets and particle accelerators.

**Introduction**

The pursuit of high-field applications, such as fusion energy and advanced particle accelerators, relies heavily on the development of high-performance superconducting materials (*1–4*). One promising candidate is the rare-earth-barium-copper-oxide (ReBCO) tape, which has shown exceptional superconducting properties, including high critical current densities and excellent thermal stability. ReBCO tapes are particularly well-suited for high-field applications due to their high transport current at high magnetic fields, as well as the fact that they do not require any post-winding heat treatment.

While the magnetic properties of ReBCO strands are well known at this point, the magnetization behavior of ReBCO cables, especially at high fields, are much less so. Cables are often required or at least desirable in large magnet devices, both to limit magnet inductance and to provide redundancy and possibly current sharing. However, magnetization measurements of cables are less common because these are larger samples, and thus different techniques are required to measure them accurately.

Our team previously measured Conductors on Round Core (CORC) wires using a DC susceptometer approach, which allows for the measurement of sufficiently large samples such that cables can be measured (*5*). Here we focus instead on simple stacks of tapes, in the configuration of tape stack cables. Our main finding is that the stacks of tapes (tape stack cables) act like thick superconductor slabs (*6–13*), with penetration fields substantially larger than those of single tapes, and, more importantly, we observed flux jump instabilities (an unwanted phenomenon) up to high magnetic fields, in this case up to 17 T. Such instabilities may be problematic for large magnets, either in terms of transport instability or field quality. We did find that slightly separating the tapes with spacers, either copper spacers or G10 (fiberglass epoxy composite) spacers, markedly reduced the tendency to flux jump. These observations should be of interest in the design of magnets for use in ReBCO based, high-field particle accelerators, magnets, and fusion reactors.

**Experimental**

A. Samples

All samples measured in this work were assembled from tape segments cut from a length of SuperPower SCS-4050 HTS tape. The tape was 4 mm wide and had a substrate thickness of 30 μm, a Cu plating thickness of 10 μm, and a tape $I_c$ of 100 A at 77 K and self-field. Three samples were measured; the first consisted of a 2.7 cm long stack of 60-tapes, the second sample was the same except that thin Cu spacers (0.16 mm thick) were placed between each tape and its neighbor, and the third sample, again 2.7 cm long, consisted of a 30-tape stack with G10 spacers (0.38 mm thick) in between each ReBCO tape and its neighbor. We used a diamond saw to cut the samples. In all cases, the field was perpendicular to the tape stack cable width, as shown in Figure 1. *M-H* Measurements on these samples were made with a magnetic field orthogonal to the cable's longitudinal axis in a LHe environment (4.2 K). The specifications of these tape stack cables are listed in Table 1.

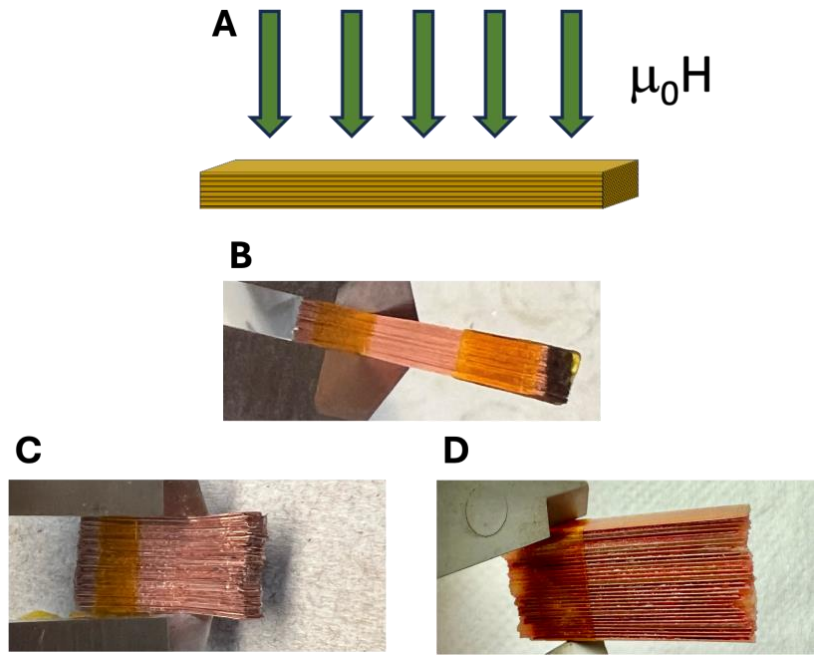

**Fig. 1. (A) Schematics of the applied field perpendicular to the tape stack cables, (B) 60-tape stack cable, (C) 60-tape stack cable with Cu spacers, (D) 30-tape stack cable with G10 spacers.**

### Table 1. Tape Sample Details

| | |
|---|---|
| Tape thickness | ~ 0.06 mm |
| Cu stabilizer thickness | 10 µm |
| ReBCO thickness | 1 µm |
| Tape width | 4 mm |
| Substrate thickness | 30 µm |
| Sample length | 27 mm |
| Critical Current @77K and self-Field, $I_c$ | 100 A |
| 60 tape stack (cable) sample thickness | 3.90 mm |
| 60 tape stack (cable) with Cu-spacers tape sample thickness | 13.80 mm |
| 30 tape stack (cable) with G10-spacers tape sample thickness | 14.00 mm |

B. Measurement Procedure

The magnetic field was supplied by a 30 T resistive magnet at the National High Magnetic Field Laboratory, Tallahassee, Florida. This magnet not only provided the applied field but also served as the primary coil for a susceptibility approach to magnetization measurement. We used our own custom-made coil set for both the pickup and compensation coils; the same setup was used in our previous work (*5*).

During the measurement, a voltage was induced in the pick-up coil (which contained the sample) by the ramping of the magnetic field produced by the resistive magnet. The induced voltage, $V$, in the coil was calculated using Faraday's law. We connected a compensation coil (the same size as the pick-up coil but with no sample inside) in anti-series with the pickup coil and measured the resulting voltage. In the absence of a sample inserted in the secondary, this nullifies the induced voltage. When a sample is placed in the pick-up coil, it disrupts the balance, causing a resultant voltage which can be integrated over time to determine the magnetic flux. This flux is then calibrated to obtain the magnetization, $M$, of the sample.

Calibration was accomplished by determining the demagnetization factor for the stacked tape cables, utilizing a well-established method (*14–17*).

**Results**

*60-Tape Stack Cable (Sample A)*

Magnetization measurements on sample A, the 60-tape stack, were performed at 4.2 K using field sweeps of ±10 T, ±20 T, and ±25 T, with a consistent ramp rate of 10 T/min. Our results are shown in Figure 2, normalized to cable volume (in this case, the same as the volume of tapes).

The full penetration field, $B_p$, for sample A (60 tapes, no spacers) was difficult to determine because of substantial flux jumping in the relevant region, although if we translate the initial curve for some of the measurements, we might estimate it as 4 T. Flux jumps are clearly quite strong about the origin, but notably persist up 17 T. Initially the retention of flux jumping to these high fields was a surprise since this is not typically reported for ReBCO tapes, although see (*18*). However, it can be remembered that $B_p$ is proportional to sample thickness, rather than width, for thin single (stand-alone) coated conductors, and thus $B_p$ is much lower than it would be expected for superconducting slabs with the same $J_c$, but much greater for thick ReBCO stacks. For thicker samples, it can be remembered that $B_p \propto J_c w$, where $w$ is the tape width. In this case, then, the magnetization loop height, $\Delta M$ will be much larger for the slabs than for the tapes, and because the tendency to flux jumping is proportional to the energy of flux motion during the thermomagnetic instability, which is $\propto \Delta M$, and thus it has been observed that flux jumping has an onset at a particular value of $\Delta M$.

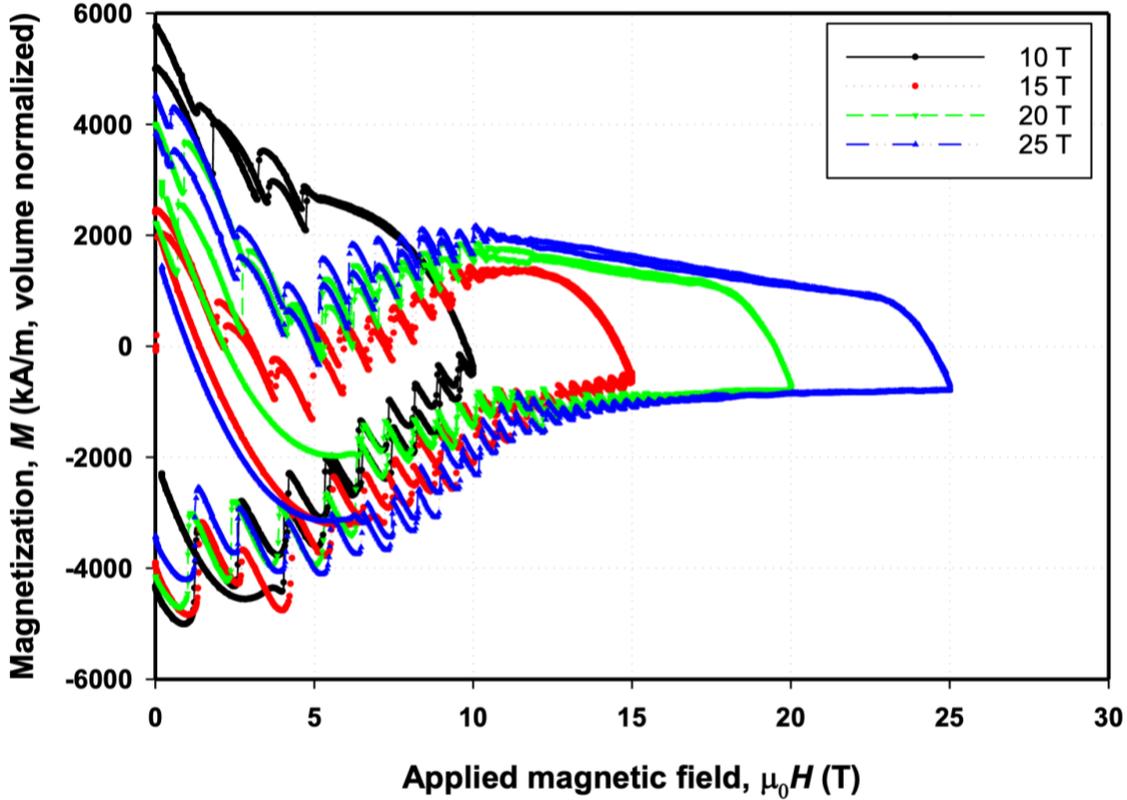

**Fig. 2. Strong flux jumping is observed up to 17 T in the *M-H* of a 60-tape stack cable measured at 4.2 K, with a maximum applied magnetic field of 25 T perpendicular to the cable width and its longitudinal axis (sample A). Magnetization is normalized to the total volume of the cable.**

Several authors have shown that stacks of tapes behave somewhat similarly to thicker slabs of superconductor, and that, in particular, $B_p$ grows with the number of tapes in a stack (*6–8, 19*). Essentially, when we stack many tapes together, the material acts like a slab conductor (with a $J_c$ reduced by $\lambda$), and this tends to lead (9) to $B_{p,\,slab} = \mu_0 \lambda J_c w/2 = \mu_0 I_c/(2 t_{tape})$. If we assume a 10 X increase in $J_c$ when reducing to 4.2 K, this leads to a $B_{p,\,slab} \cong 6$ T in the limit of a thick slab, which is much larger than would be found for an isolated tape, given by $B_{p,\,tape} = (5/2\pi)B_d[\mathrm{Ln}(w/t_{YBCO}+1)]$, where $B_d = 0.4\mu_0 I_c/w \cong 0.126$ T, such that $B_{p,tape} \cong 0.9$ T at 4.2 K. We might then expect a potential enhancement of flux jumping in stacks of ReBCO tapes for fields below the penetration field of the stack, namely $B_{p,\,slab} \cong 6$ T in our case. But in fact, we see flux jumping up to 17 T.

*60-Tape Stack Cable with Cu spacers (Sample B)*

A set of magnetization measurements were conducted on Sample B, a 60-tape stack cable incorporating Cu spacers (60 ReBCO tapes, 61 Cu spacers, stacked as Cu/ReBCO/Cu/ReBCO/Cu, etc.). As before, *M-H* measurements were performed at 4.2 K, employing field sweeps of ±10 T, ±15 T, ±20 T, ±25 T, and ±30 T, with a consistent ramp rate of 10 T/min. The resulting magnetization data, normalized to total cable volume (ReBCO tapes + spacers), are presented in Figure 3.

Sample B shows a markedly lower tendency to flux jump as compared to sample A. Here we can observe a $B_p$ = 2.5 T and a corresponding magnetization, $M_p$, of 1350 kA/m. Flux jumps are seen only up to an applied field of 5 T, and they are much smaller in size than those of Sample A. We also note that little flux jumping occurs for *M* less than 1000 kA/m.

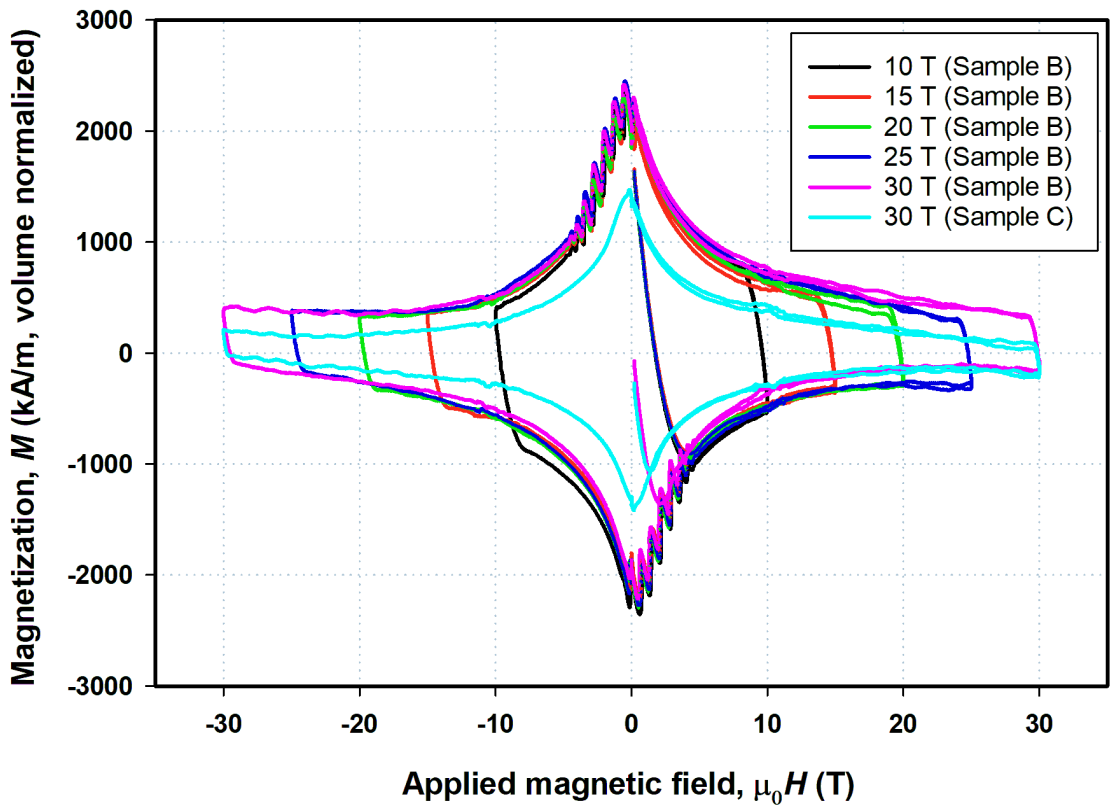

**Fig. 3. The *M-H* of sample B, a 60-tape stack cable with Cu spacers (marked as 10 T, 15 T, 20 T, 25 T, 30 T) and Sample C, a 30-tape stack cable with G10 spacers (marked as G10), each measured at 4.2 K, with a maximum applied magnetic field of 30 T perpendicular to the longitudinal axis of the cable and to its width. Notice the large suppression of flux jump events. Magnetization is normalized to the total volume of the cable.**

*30-Tape Stack Cable with G10 spacers (Sample C)*

A set of magnetization measurements were conducted on sample C, a 30-tape stack cable incorporating G10 spacers (30 ReBCO tapes, 31 G10 spacers). As before, *M-H* measurements were performed at 4.2 K, employing field sweeps of ±10 T, ±15 T, ±20 T, ±25 T, and ±30 T, with a consistent ramp rate of 10 T/min. The resulting magnetization data, normalized to cable volume, are presented in Figure 3 for the 30T run only, along with the data from Sample B. The results for lower amplitude sweeps were very similar (omitted for clarity).

Sample C exhibited a full penetration field, $B_p$ = 1.5 T, and a corresponding magnetization, $M_p$, of 1050 kA/m, indicative of full flux penetration. No flux jumping was observed in this sample.

**Discussion**

Using the "dilute" superconductor model (*20*), the penetration field for sample A was estimated to be $B_{p,A}$ = 6.3 T, while that for Sample B can be estimated (including the spacer thickness in the composite volume) as $B_{p,B}$ = 2.86 T. Using the same approach for Sample C we find $B_{p,C}$ = 1.4 T. The experimentally observed full penetration field is 4 T (extrapolated) for Sample A, and 2.5 T and 1.5 T for Samples B and C, respectively. The maximum magnetization of Sample A, namely 5000 kA/m, is reduced to 2400 kA/m for Sample B and to 1400 kA/m for Sample C once we take the whole composite volume (including spacers) into account. We observe in all samples reduced flux jumping below magnetization values of 1000 kA/m. The fact that separating the tapes slightly with spacers reduces flux jumping below the modified penetration fields of these conductors is not surprising. However, the suppression of flux jumping in samples with spacers persists to much higher fields, and the flux jumping onset (as we move from high fields to low ones) appears to correlate with a fixed value of composite magnetization, which is interesting and suggests that some dynamic cooling is present.

**Conclusion**

This work explores the magnetization of stacks of ReBCO tapes in magnetic fields up to 30 T in the configuration of a tape stack cable. In some samples, flux jumping was observed to persist up to 17 T. Such effects are undesirable and can lead to unwanted signatures in transport properties and in the degradation of magnet field quality. In this work, the applied magnetic fields were perpendicular to the flat face of the tapes, and all measurements were performed at 4.2 K, with a ramp rate of 10 T/min. Magnetic fields were applied using a resistive magnet at the NMHFL in Tallahassee, and we employed a DC susceptometer approach with a compensator coil to measure the *M-H* curve of the samples. In our first sample (Sample A), which was a simple stack of 60 tapes, we observed both an increase in the penetration field (compared to the single tape) and the presence of flux jumping up to 17 T. A second sample (Sample B) was made with Cu spacers in between each ReBCO tape, and strong suppression of flux jumping was observed. A third sample (Sample C) was made with further increased inter-tape spacing, in this case with G10 spacers, and flux jumping was removed entirely. These results should be of substantial interest to the magnet community, and in particular, those involved with either accelerator or fusion-related HTS magnets.

**Acknowledgments:** This work was supported in part by the United States Department of Energy, Office of Science, Division of High Energy Physics under Grant DE-SC0011721, in part by the National Science Foundation Cooperative under Agreement DMR-2128556 which was performed at the National High Magnetic Field Laboratory, and in part by the State of Florida.

**Funding:** Provide complete funding information, including grant numbers, complete funding agency names, and recipient's initials. Each funding source should be listed in a separate paragraph.

10.13039/100000015 – U.S. Department of Energy, Office of Science, Division of High Energy Physics (Grant Number: DE-SC0011721) (MDS, TG, EWC, MM)

National Science Foundation Cooperative (Grant Number: DMR-2128556) (JJ, ESC)

10.13039/100012892 – National High Magnetic Field Laboratory (JJ, ESC)

10.13039/100023043 – State of Florida (JJ, ESC)

**Author contributions:**

Conceptualization: TG, MDS

Methodology: TG, MDS, JJ, ESC

Instrumentation: TG, JJ, ESC

Investigation: TG, MDS

Visualization: MDS, MM

Funding acquisition: MDS

Project administration: MDS

Supervision: MDS, EWC

Writing – original draft: TG, MDS

Writing – review & editing: TG, MDS, EWC

**Competing interests:** The authors declare that they have no competing interests.

**Data and materials availability:** All data are available in the main text or the supplementary materials.